\documentclass[aps,floats,twocolumn,superscriptaddress]{revtex4}
\usepackage[final]{graphicx}
\DeclareGraphicsExtensions{.eps,.ps,.pdf}
\usepackage{amsmath,amssymb,bbold,bm}
\def\id{\mathbb{1}}
\def\tr{\mathrm{tr}}

\def\ii{\mathrm{i}}

\newcommand{\bk}{{\bf k}}
\newcommand{\bp}{{\bf p}}

\newcommand{\br}{{\bf r}}

\newcommand{\bB}{{\bf B}}
\newcommand{\bA}{{\bf A}}
\newcommand{\bE}{{\bf E}}

\begin{document}

\title{Exciton condensation and charge fractionalization in a topological insulator film}
\author{B. Seradjeh}
\affiliation{Department of Physics,
University of Illinois, 1110 West Green St, Urbana, IL 61801-3080}
\author{J.E. Moore}
\affiliation{Department of Physics, University of California, Berkeley, CA 94720}
\affiliation{Materials Sciences Division, Lawrence Berkeley National Laboratory, Berkeley, CA 94720}
\author{M. Franz}
\affiliation{Department of Physics and Astronomy, University of British Columbia,Vancouver, BC, Canada V6T 1Z1}
\date{\today}

\begin{abstract}
  An odd number of gapless Dirac fermions is guaranteed to exist
  at a surface of a strong topological insulator. We show that in a
  thin-film geometry and under external bias, electron-hole pairs that
  reside in these surface states can condense to form a novel exotic quantum state 
which we propose to call `topological exciton condensate' (TEC). This TEC is
 similar in general terms to the exciton condensate recently argued to exist in a biased graphene bilayer, but with different topological properties. It  exhibits a host of unusual properties including a stable zero mode and a fractional charge $\pm e/2$ carried by
  a singly quantized vortex in the TEC order parameter.
\end{abstract}

\maketitle

\emph{Introduction}.---Recent advances in studies of band insulators
with strong spin-orbit coupling revealed the existence of new
topological invariants that characterize these
materials~\cite{mele1}. 
Among the three-dimensional
time-reversal (${\cal T}$) invariant insulators, the most interesting
phase implied by this classification is the ``strong'' topological
insulator (STI), characterized by gapless fermionic states residing at
its surface with an {\em odd} number of topologically protected
nodes. These gapless states exhibit linear dispersion and behave as
massless Dirac fermions familiar from the physics of graphene. Several
real materials have been identified as STIs in pioneering
experiments~\cite{cava1,cava2} completed shortly after the theoretical
predictions~\cite{kane1}. These rapid developments give hope that the
new state of quantum matter realized in STIs might be relatively
common in nature and raise the prospects of future practical
applications.

The existence of an odd number of Dirac fermions leads to a number of
exotic properties associated with surfaces of a STI. These include an
exotic superconducting state induced by a proximity effect that
supports Majorana fermions~\cite{fu1}, a ${\cal T}$-breaking phase which exhibits a {\em fractional} quantum Hall effect~\cite{qi1}, and an unusual
`axion' electromagnetic response~\cite{qi1,essin1}.

\begin{figure}[t]
\includegraphics[width=7cm]{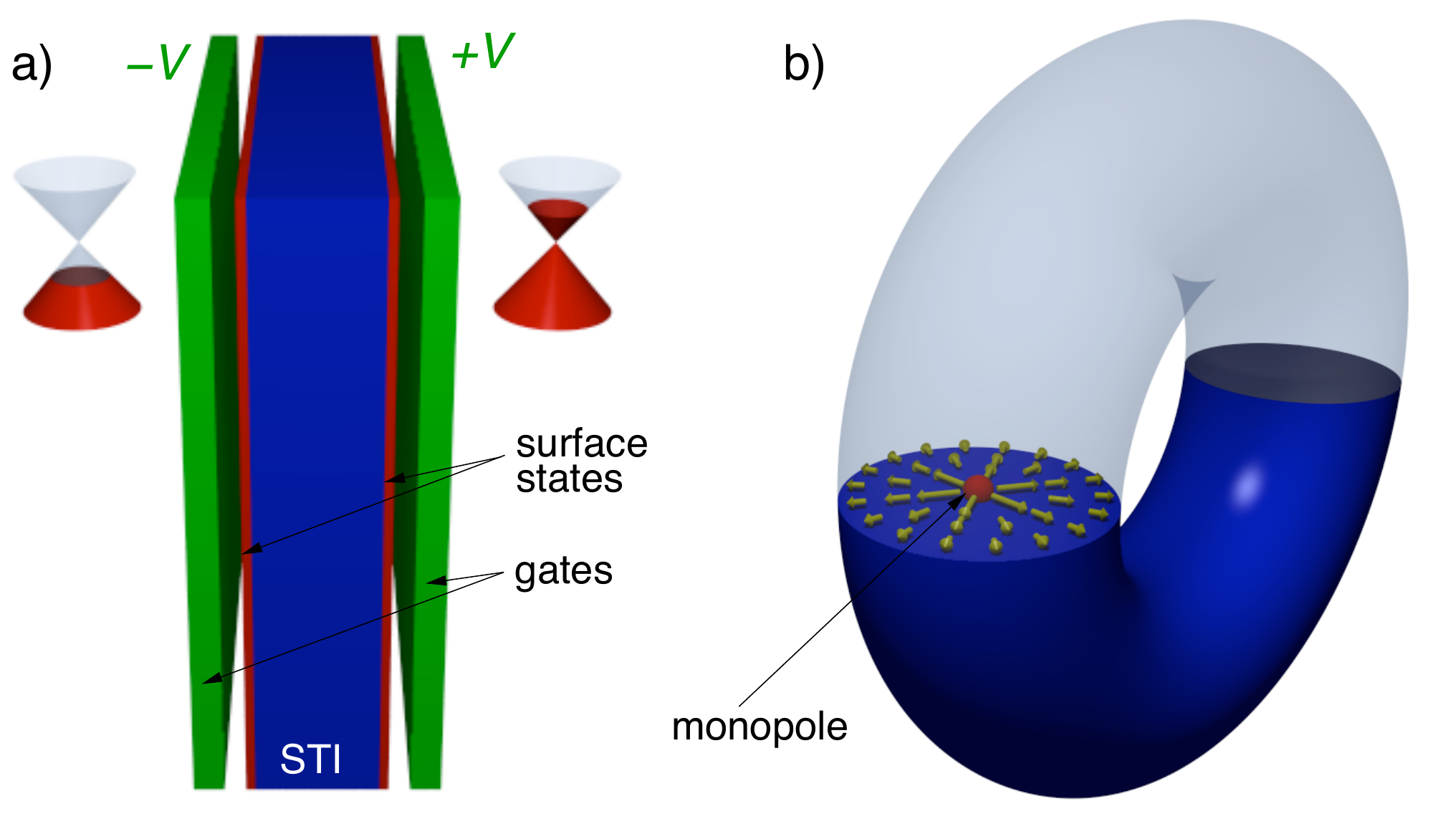}
\caption{(Color online) a) Schematic of the proposed device. b) The
  exciton condensate effectively joins the surfaces of the STI film
  resulting in toroidal topology. Arrows illustrate the magnetic
  field distribution of a planar monopole representing a
  vortex in the effective theory.}
\label{fig1}
\end{figure}
The wealth of exotic phenomena listed above stems from the possibility
of inducing various types of mass terms in the otherwise massless Dirac
fermion states at the surface of a STI. In this Letter we introduce and study
a new type of mass gap that can be induced by a Coulomb interaction between 
the surface states of a thin STI film and can be characterized as a 
`topological' exciton condensate (TEC). The idea is motivated by recent
proposals to realize an exciton condensate in a symmetrically biased graphene
bilayer~\cite{lozovik1,min1}. We argue below that TEC in an STI film might be
more easily realized than in graphene and is a different, genuinely topological phase,
distinguished by the presence of a zero-energy mode and fractional charge associated with its vortices.

Consider a film made of a 
STI placed inside a
capacitor as in Fig.\ \ref{fig1}a. Imagine for simplicity that each
surface harbors a single Dirac cone with the chemical potential $\mu$
initially tuned to the neutral point $\mu=0$. When the
capacitor is charged the Fermi levels in the two layers move in the
opposite direction, creating a small electron Fermi surface in one
layer and a small hole Fermi surface in the other. For arbitrarily weak
repulsive interaction such a system will form an exciton condensate
which may be pictured as a coherent liquid of electron-hole pairs
residing in different layers.

In what follows we use a simple model for the surface states to show
how exciton condensation can be induced by the interlayer Coulomb
interaction. By examining this model we then deduce some
interesting properties of the underlying TEC. Specifically,
we demonstrate that an isolated singly quantized vortex in the complex
scalar order parameter characterizing TEC contains a zero mode and carries
topologically protected exact fractional charge $\pm e/2$. 
We put our findings in the context of axion electrodynamics, the
low-energy effective theory of STIs, and discuss prospects for experimental
 realization and detection of the predicted phenomena.

\emph{Model}.---The gapless states associated with the two surfaces
of the biased STI film can be described at low energies by a Dirac
Hamiltonian~\cite{fu1}
\begin{equation}\label{h0}
H=\sum_{l=1,2}\psi^\dagger_l(v_l\bm{\sigma}\cdot\hat\bp-\mu_l)\psi_l+Un_1n_2,
\end{equation}
where $\psi_l=(c_{l\uparrow},c_{l\downarrow})^T$ denotes the fermion
spinor in surface layer $l$, $ \bm{\sigma}=(\sigma_x,\sigma_y)$ is
the vector of Pauli matrices in the spin space, $\hat\bp=-i\nabla$,
and $v_l=(-1)^{l+1}v$ represents the Fermi velocity, assumed to be
opposite at the two surfaces and
$n_l=\psi^\dagger_l\psi_l$. We take $\mu_l=\mu+(-1)^lV$, with $\mu$
the intrinsic chemical potential and $V$ the external bias. The last
term in (\ref{h0}) describes the short-range part of the interlayer
Coulomb potential, a sufficient minimal interaction for the formation of
 EC~\cite{ser1}. Also, we assume the film to be sufficiently thick so that any
direct hopping of low-energy electrons between the surfaces can be neglected.

To describe the exciton condensation we decouple the interaction term
in $H$ using a matrix-valued order parameter $M=U\langle
\psi_1\psi^\dagger_2\rangle$. The expectation value is taken with
respect to the mean-field Hamiltonian
\begin{equation}\label{h1} H_{\rm MF}=H_0+(\psi^\dagger_1M\psi_2 +
{\rm h.c.})  +{1\over U}{\rm Tr}(M^\dagger M),
\end{equation}
where $H_0$ denotes the kinetic term in (\ref{h0}). At this point it
is useful to organize the Fermi fields into a single 4-component
spinor $\Psi=(\psi_1,\psi_2)^T$. We can write $H_{\rm MF}=\Psi^\dagger
{\cal H}\Psi +\case{1}{U}{\rm Tr}(M^\dagger M)$ with a $4\times 4$
matrix Hamiltonian
\begin{equation}\label{h2} {\cal H}+\mu= \left(
\begin{array}{cc} v\bm{\sigma}\cdot\hat\bp-V & M \\ M^\dagger &
-v\bm{\sigma}\cdot\hat\bp+V
\end{array} \right).
\end{equation}

Various forms of matrix $M$ describe different possibilities for the
TEC order parameter.  When $\mu$ is close to
zero an order parameter that opens up a gap in the excitation spectrum
will be favored because it leads to an overall reduction in kinetic
energy. In the uniform system this will occur only for $M$ diagonal in
the spin space, i.e.\ $M=m\mathbb{1}$ with $m$ a complex constant,
since the part of ${\cal H}$ proportional to $m$ then anticommutes
with the kinetic term. The spectrum of ${\cal H}$ then contains four
branches and reads
\begin{equation}\label{spec}
E_{\bk\alpha s}=-\mu+\alpha\sqrt{(v|\bk|+sV)^2+|m|^2}, \ \
\alpha,s=\pm 1.
\end{equation}
Physically, this form of the matrix $M$ implies non-zero expectation
values $\langle c_{1\uparrow}c^\dagger_{2\uparrow}\rangle=\langle
c_{1\downarrow}c^\dagger_{2\downarrow}\rangle=m/U$. We also note that
this is the only choice of the order parameter that leaves the Hamiltonian ${\cal
T}$-invariant.

\emph{Exciton condensate}.---When $\mu=0$ and $V\neq 0$ the exciton
instability of our Hamiltonian (\ref{h0}) is formally equivalent to
the Cooper instability in a metal and occurs at infinitesimal coupling
$U$. To see what happens when $\mu$ is slightly detuned from zero, we
consider the gap equation for $m$ (now assumed to be real) obtained by
minimizing the ground state energy
$E_g=\sum_{\bk\alpha s}'E_{\bk\alpha s}+Nm^2/U$ with respect to
$m$. This reads
\begin{equation}\label{gap} {m\over U}=-{1\over
2N}{\sum_{\bk\alpha s}}'{m\over E_{\bk\alpha s}+\mu},
\end{equation}
where the prime denotes a sum over the occupied states
($E_{\bk\alpha s}<0$) only. For general values of $\mu$, $V$ and $U$ the
gap equation must be solved numerically. One can, however, extract the
value $U_c$ of the critical coupling beyond which TEC is formed. In
the experimentally relevant regime $\mu\ll V\ll\Lambda$, where
$\Lambda$ denotes the high-energy cutoff of the order of bandwidth, we
find
\begin{equation}\label{uc} U_c\simeq {4\Lambda} \left(1+{V\over
\Lambda}\ln{V\over\mu}\right)^{-1}.
\end{equation}
The critical coupling is large, of the order of bandwidth, unless
$\mu\ll V e^{-\Lambda/V}$, in which case $U_c$ becomes small and
eventually reaches zero when $\mu\to 0$. In this limit the gap
equation can be solved explicitly to obtain $m\approx
2\sqrt{V\Lambda}e^{-\Lambda^2/UV}$. In order to achieve exciton
condensation for a given coupling strength $U$ it is essential to tune
$\mu$ as close to zero as possible and apply high bias voltage $V$.
Henceforth we consider only the $\mu=0$ situation.

Because of its exponential dependence on the coupling strength it is
difficult to give a truly quantitative estimate of $m$ and the
relevant TEC transition temperature $T_c$ for a realistic STI film. In
the context of the graphene bilayer the estimates of $T_c$
range from sub-Kelvin up to the room temperature, depending
on the approximation employed~\cite{lozovik1,min1,efetov1}. 
Although we do not attempt such a
quantitative analysis here we note that the situation in STI might be
quite similar. On the one hand the intrinsic
energy scales in known STIs are somewhat smaller than in graphene. On
the other hand, screening is known to reduce the mean-field
$T_c$ by a factor $\sim e^{\cal N}$, where ${\cal N}$ denotes the
number of surface Dirac modes. Formation of TEC in an STI film made from
Bi$_2$Se$_3$ will be therefore aided by the fact that this material exhibits
${\cal N}=1$~\cite{cava1} compared to ${\cal N}=4$ in graphene (due to
valley and spin degeneracies).

\emph{Vortex zero modes}.---In the following we adopt the point of view
that, based on the above analysis, formation of TEC is likely to occur
under experimentally achievable conditions and focus on its unique
properties. To this end it is convenient to write the Hamiltonian
(\ref{h2}) in a more customary form using the Dirac matrices in the
Weyl representation, $\gamma_j=i\tau_2\otimes\sigma_j$, $j=1,2,3$,
$\gamma_0=\tau_1\otimes\id$ and
$\gamma_5=-i\gamma_0\gamma_1\gamma_2\gamma_3=\tau_3\otimes\id$, where
$\tau_j$ are Pauli matrices in the layer space. We obtain
\begin{equation}\label{h3} {\cal H}=\gamma_0\left(\gamma_1\hat
p_x+\gamma_2\hat p_y+V\gamma_0\gamma_5 + |m|e^{-\ii\gamma_5\chi}
\right),
\end{equation}
where we have set $v=1$ and used a polar representation $m=|m|e^{i\chi}$ of the
complex TEC order parameter.

The Hamiltonian (\ref{h3}) formally coincides with the one used to describe
the effect of EC near one of the valleys of the biased graphene bilayer
system~\cite{ser1}. That work established existence of an {\em exact zero
mode} of the Hamiltonian (\ref{h3}) in the presence of a singly
quantized vortex in the EC order parameter $m=m_0e^{i\varphi}$, with
$\varphi$ the polar angle. The zero-energy eigenstate has the form
$\Psi_0=(f,g,ig^*,-if^*)^T$ with $f=Ae^{-m_0r}J_0(Vr)$,
$g=-iAe^{i\theta}e^{-m_0r}J_1(Vr)$ and $A$ the normalization constant.
In graphene, valleys always come in pairs. The fermionic zero modes
are thus doubled and split due to intervalley scattering. Because of
this, no exact zero modes survive in the graphene bilayer.

In a STI, by contrast, there is always an {\em odd number} of valleys
associated with the surface~\cite{mele1}.
When ${\cal N}=1$, as in Bi$_2$Se$_3$, the zero mode attached to a singly
quantized vortex will remain exact, as long as the vortex stays well
separated from other vortices or system edges. This finding, along
with the fractional charge discussed below, constitutes the key universal difference between the STI exciton condensate and other proposed exciton condensates and is the main result of this work.

For general odd ${\cal N}>1$ we expect ${\cal N}-1$ zero modes to
split symmetrically around zero energy while the remaining zero mode
will persist. This conclusion follows from the property $\gamma_2
{\cal H}^*\gamma_2={\cal H}$ which together with
$\gamma_2^2=-1$ implies the spectral symmetry around the zero energy
of eigenstates of ${\cal H}$. Specifically, for each eigenstate
$\Psi_E$ of energy $E$ there exists an eigenstate $\Omega\Psi_E$ with
energy $-E$. Here $\Omega=\gamma_2K$ is an antiunitary operator and
$K$ denotes complex conjugation. It also holds that $\{\Omega,{\cal
H}\}=0$.

\emph{Fractional charge}.---A localized zero mode in a particle-hole
symmetric system is known to carry a fractional charge $\pm e/2$~\cite{JacReb76a,SuSchHee79a,GolWil81a,HouChaMud07a}. Thus, we expect
our vortices to be fractionally charged. To see how this occurs in the
present system and also to deduce some of its other interesting
properties, let us consider an operator $O$, represented by a constant
$4\times 4$ Hermitian matrix, acting in the space of wavefunctions
$\Psi_E(\br)$. Following~\cite{herbut1} consider now the quantity
\begin{equation}\label{c1} {\cal
O}\equiv\sum_E\langle\Psi_E|O|\Psi_E\rangle = {1\over 4}N\tr(O),
\end{equation}
where $N$ is the total number of quantum states in a suitably
regularized system (e.g.\ on the lattice and in finite volume).  The
last equality in (\ref{c1}) follows from the completeness of states.
We may write
\begin{equation}\label{c2} {\cal
O}=\left(\sum_{E<0}+\sum_{E>0}\right)\langle\Psi_E|O|\Psi_E\rangle +
\langle\Psi_0|O|\Psi_0\rangle
\end{equation}
The spectral symmetry generated by $\Omega$ and its antiunitarity,
expressed as $\langle\Omega\Psi_1|\Omega\Psi_2\rangle=
\langle\Psi_1|\Psi_2\rangle^*$, imply
$\label{c3} \sum_{E>0}\langle\Psi_E|O|\Psi_E\rangle =
\sum_{E<0}\langle\Psi_E|(\Omega^{-1}O\Omega)^\dagger|\Psi_E\rangle.$ 
If $O$ furthermore {\em commutes} with $\Omega$ then the last term
becomes simply $\sum_{E<0}\langle\Psi_E|O|\Psi_E\rangle$ and we can
combine Eqs.\ (\ref{c1}-\ref{c2}) to obtain
\begin{equation}\label{c4} \sum_{E<0}\langle\Psi_E|O|\Psi_E\rangle =
{1\over 2}\left[{N\over 4}\tr(O)-
\langle\Psi_0|O|\Psi_0\rangle\right].
\end{equation}
The expectation value of an observable
represented by a constant $4\times 4$ matrix that commutes with
$\Omega$, taken over all occupied negative-energy eigenstates of
${\cal H}$, is determined solely by the value of $\tr(O)$ and the
zero-mode eigenstate of ${\cal H}$. For an infinite system in
continuum Eq.\ (\ref{c4}) will be useful for quantities independent of
$N$; this occurs when $O$ is traceless or else for quantities
represented as differences so that $N\tr(O)$ drops out. In such cases
the expectation value only depends on the zero mode and its value is
expected to be robust. Specific examples follow below.

The charge operator is represented by a $4\times 4$ unit matrix
$O_Q=e\id$. The charge bound to a vortex can be expressed as
\begin{equation}\label{c5} Q_V=
e\sum_{E<0}\bigl(\langle\Psi_E|\id|\Psi_E\rangle_1
-\langle\Psi_E|\id|\Psi_E\rangle_0\bigr),
\end{equation}
where subscripts $1$ and $0$ refer to the state with one and zero
vortices respectively. Using Eq.\ (\ref{c4}) we find
$ Q_V= -{e\over 2}\langle\Psi_0|\Psi_0\rangle_1=-{e\over 2},$
as expected. We note that Eq.\ (\ref{c5}) assumed the zero mode to be
unoccupied; if we occupy it by an electron then the vortex charge becomes
$+e/2$.

Other quantities of interest include the spin ${\bf S}$ and the axial
charge $Q^5$ carried by the vortex, defined as the charge
difference between the layers. These are represented by matrices
$O_{\bf S}=\frac{1}{2}\gamma_0\vec{\gamma}\gamma_5$ and
$O_{Q^5}=e\gamma_5$. Unfortunately these anticommute with $\Omega$
making Eq.\ (\ref{c4}) inapplicable. A quantity that can be calculated
is the interlayer spin polarization $\Delta{\bf S}$, represented by
$O_{\Delta\bf S}=\frac{1}{2}\gamma_0\vec{\gamma}$. A straightforward
calculation shows that, in the presence of a vortex, $\langle\Delta
S_x\rangle=\langle\Delta S_y\rangle=0$ while $\langle\Delta
S_z\rangle$ varies smoothly from ${1\over 2}$ to $0$ as we tune $V/m$
from 0 to infinity. This implies that the vortex carries a fractional
value of spin polarization between the layers. We note, however, that
due to the entanglement of the spin and momenta in the STI, the total
spin operators in the TEC or the STI surfaces are not sharp.

The symmetry generated by $\Omega$ is a combination of ${\cal T}$ and
spatial parity ${\cal P}$. The latter will be broken in the presence
of non-magnetic impurities and the zero mode will no longer be
exact. However, the following general argument shows that the
fractional charge remains precisely quantized as long as the bulk is
gapped and ${\cal T}$ is preserved.

\emph{Axion electrodynamics}.--- As demonstrated in
Refs.~\cite{qi1,essin1} the response of STI to external
electromagnetic
field is that of an `axion' medium~\cite{wilczek1} and can be
mathematically implemented by adding a term
\begin{equation}\label{ax1} \Delta{\cal L}_{\rm axion}=\frac\theta{2\pi}
\frac{e^2}{hc}\:\bB\cdot\bE
\end{equation}
to the usual Maxwell Lagrangian. An ordinary insulator has $\theta=0$
while the STI exhibits $\theta=\pi$, the two values permitted by the
time-reversal symmetry. When a surface of a STI is gapped by a ${\cal
T}$-breaking perturbation, such as an applied magnetic field, $\theta$
varies smoothly between $\pi$ and 0.  As a result the surface behaves
as a quantum Hall fluid~\cite{wilczek1,qi1}.

As noted above, TEC does not break ${\cal T}$. Instead,  the
TEC order parameter effectively identifies the opposite surfaces of
the film by allowing, at the mean-field level of the Hamiltonian
(\ref{h1}), the electrons to hop between them. Dirac fermions at the surfaces 
are gapped without violating ${\cal T}$ by a perturbation that effectively 
removes the surfaces as illustrated in Fig.\ \ref{fig1}b.
Because the
TEC order parameter is in general complex the electrons hopping
between the surfaces  may acquire a nontrivial phase. This is easily
included by postulating twisted periodic boundary conditions with a
twist equal to the phase of the TEC order parameter $\chi$. The
situation is easiest to visualize in the context of a lattice model of
STI where electrons traversing the bonds that connect the two surfaces
acquire a phase $\chi$. This extra phase can be also viewed as
resulting from an electromagnetic vector potential
$\delta\bA=\hat{z}\chi(\Phi_0/a)$ localized in the layer between the
surfaces, with $\Phi_0=hc/e$ the flux quantum and $a$ the lattice
spacing.

In a vortex configuration with
$\chi(\br)=\varphi$ it is easy to see that $\delta\bA(\br)$ has the form of a
planar {\em monopole}: the magnetic field $\delta\bB$ radiates outward
from the vortex center in the plane of the surface as illustrated in
Fig.\ \ref{fig1}b. The total flux is $\Phi_0$. This vector potential is
fictitious in the sense that $\delta\bB$ cannot be detected by an
outside probe. To electrons, however, $\delta\bA$ is indistinguishable
from the real vector potential and must be included in Eq.\
(\ref{ax1}) when evaluating the response of the STI.
A unit magnetic monopole in a `$\theta$-vacuum' described by Eq.\
(\ref{ax1}) is known to carry electric charge $-e(\theta/2\pi+n)$ with
$n$ integer~\cite{witten1}. Applied to TEC this gives vortex charge
$-e(1/2+n)$, consistent with our finding of the fractional charge $\pm
e/2$ bound to the vortex.

\emph{Outlook and open questions}.--- Recent advances in materials
engineering and fabrication give hope that the exotic state of matter
identified in this work can be achieved and probed in the near
future. From an experimental point of view, there are several
significant advantages of the proposed phase. First, we believe that
making an exciton condensate between the surfaces of a single film,
rather than from two 2D materials (e.g., graphene) with an insulator
between them, is easier because it does not require creating a
pinhole-free insulating layer with defect-free junctions to the two 2D
materials.  Second, attaching leads to a surface of a film should be
considerably easier than attaching leads to graphene. Once the leads
are in place it should be straightforward to identify the onset of the
exciton condensation in a transport measurement as demonstrated in
recent literature~\cite{eisen1}. Third,  we note that the existence of the
zero-energy Dirac fermion can be detected in the same way as a
standard midgap impurity state by optical methods or in a careful
transport measurement.  In the TEC phase a transport measurement will
reveal a gap at low temperature. As the temperature nears the
transition temperature, conduction becomes dominated by the charges
bound to vortices and is proportional to the number of thermally
excited vortices. Fractional charge can be probed, at least in principle, by the shot noise analysis of resistivity ~\cite{saminadayar1}.

From a theoretical point of view TEC in STI film is fundamentally
interesting for several reasons.  To our knowledge, TEC is the first
example of a new symmetry-breaking phase enabled by the special
properties of topological insulators.  It differs from the
superconducting state generated at the surface by proximity
effect~\cite{fu1} because in that case there is no new symmetry
breaking.  It differs from the ordinary exciton condensate, which is
in the same universality class as a $^4$He film, because the zero mode
attached to a vortex is stable. In the case of graphene bilayer the
zero modes are not protected due to inter-valley scattering.  In
this respect as in several others, topological insulators allow the
realization of physics that is spoiled in graphene by inter-valley
scattering.  The consequence is a distinct low-temperature phase of
matter with fractionally charged topological excitations whose
exchange statistics presents an interesting open question.

\emph{Acknowledgment}.---Support for this work came from NSERC,
CIfAR, NSF, ICMT at UIUC and the Killam Foundation. The authors also acknowledge The
Banff International Research Station where this collaboration was
initiated.


\end{document}